# Noise Characterization of IUCAA Digital Sampling Array Controller (IDSAC)

Sabyasachi Chattopadhyay[a,*], A. N. Ramaprakash[a], Bhushan Joshi[a], Pravin A. Chordia[a], Mahesh P. Burse[a], Kalpesh Chillal[a], Sakya Sinha[a], Sujit Punnadi[a], Ketan Rikame[b], Sungwook E. Hong[c], Dhruv Paranjpye[d], Haeun Chung[e,f], Changbom Park[e], Amitesh Omar[g]

[a]Inter-University Centre for Astronomy and Astrophysics, Pune, India
[b]Indian Institute of Science Education and Research, Pune, India
[c]Korea Astronomy and Space Science Institute, 776 Daedeokdae-ro, Yuseong-gu, Daejeon 34055, Korea
[d]Maharashtra Institute of Technology, Pune, India
[e]Korea Institute for Advanced Study, 85 Heogiro, Dongdaemun-gu, Seoul 02455, Korea
[f]Department of Physics and Astronomy, Seoul National University, 1 Gwanak-ro, Gwanak-gu, Seoul 08826, Korea
[g]Aryabhatta Research Institute of Observational Sciences, Nainital, India

**Abstract.**
IUCAA Digital Sampling Array Controller (IDSAC) is a flexible and generic yet powerful CCD controller which can handle a wide range of scientific detectors. Based on an easily scalable modular backplane architecture consisting of Single Board Controllers (SBC), IDSAC can control large detector arrays and mosaics. Each of the SBCs offers the full functionality required to control a CCD independently. The SBCs can be cold swapped without the need to reconfigure them. IDSAC is also available in a backplane-less architecture. Each SBC can handle data from up to four video channels with or without dummy outputs at speeds up to 500 kilo Pixels Per Second (kPPS) Per Channel with a resolution of 16 bits. Communication with Linux based host computer is through a USB3.0 interface, with the option of using copper or optical fibers. A Field Programmable Gate Array (FPGA) is used as the master controller in each SBC which allows great flexibility in optimizing performance by adjusting gain, timing signals, bias levels, etc. using user-editable configuration files without altering the circuit topology. Elimination of thermal kTC noise is achieved via Digital Correlated Double Sampling (DCDS). The number of digital samples per pixel (for both reset and signal levels) are user configurable. We present the results of noise performance characterization of IDSAC through simulation, theoretical modeling, and actual measurements. The contribution of different types of noise sources is modeled using a tool to predict noise of a generic DCDS signal chain analytically. The analytical model predicts the net input referenced noise of the signal chain to be 5 electrons for 200k pixels per second per channel readout rate with 3 samples per pixel. Using a cryogenic test set up in the lab, the noise is measured to be 5.4 e (24.3 $\mu$V), for the same readout configuration. With a better-optimized configuration of 500 kPPS readout rate, the measured noise is down to 3.8 electrons RMS (17 $\mu$V), with 3 samples per interval.

*Sabyasachi Chattopdhyay, sabyasachi@iucaa.in

## 1 Introduction

The dominant contributor to noise in a cooled detector is reset noise which results from the output amplifier for each pixel settling to a different reference level after every reset operation (1). Elimination of reset noise from CCD readout electronics (also called analog signal chain) is achieved by correlated double sampling (CDS - 2). In CDS, the output amplifier is measured twice, once with a fixed voltage (reference level) and once with the pixel charge (data level), during every pixel readout. The difference between the two levels accurately determines the amount of charge collected by a particular pixel while eliminating the reset noise. Further noise reduction is achieved in the analog domain (Analog CDS - ACDS) typically by using a low pass filter (an integrator) in the signal chain (3, 4).

Demands of very high image resolutions over large fields of view in modern-day astronomical applications have resulted in the need for large focal plane detector arrays (5, 6, 7, 8). Thus it is a continuing need to develop faster readout electronics with considerably lesser noise. An alternative approach of implementing CDS technique in the digital domain (DCDS) can lead to a significantly more compact signal processing chain and potentially improved noise performance. DCDS also offers flexibility in pixel readout speed and sampling rate having to alter the circuit. Several DCDS implementations exist through the works of Wu et al., Gach et al., Clapp, Obrosslak et al., Smith et al. etc (9–13).

IUCAA Digital Sampler Array Controller (IDSAC - 14) is an easily scalable and reconfigurable detector electronics and data acquisition system. Each Single Board Controller (SBC) element of IDSAC can handle up to four detector outputs with or without dummy outputs. A versatile DCDS technique is employed in each of the signal processing chains.

This paper describes the performance analysis of the IDSAC DCDS signal chain through analytical modeling, component level simulation and laboratory measurement. A separate article will describe the controller in detail (Burse et al., in preparation). For completeness, in Section 2 we will briefly present the design of IDSAC and its functional aspects. We discuss an analytical noise model of DCDS as it is implemented in IDSAC in Section 3. In Section 4 we discuss various tests which were performed to characterize IDSAC e.g. ADC noise, bias noise, signal chain noise etc. Section 5 summarizes the results.

## 2 System Level Architecture of IDSAC

The salient feature of IDSAC is its flexibility introduced by its architecture as shown in the Figure 1. The user can configure all the parameters from files without changing the firmware loaded in the Field Programmable Gate Array (FPGA). The host software runs on a Fedora 20 Linux PC (and other Linux variants also) which reads the parameter files from the host PC and sends commands to the IDSAC firmware via a USB cable. In return, the FPGA sends acknowledgments or data whichever is applicable. Based on the commands received, the FPGA configures the biases and clocks as well as read ADC data. Data received from FPGA is then reformatted by the PC software to create FITS formatted images.

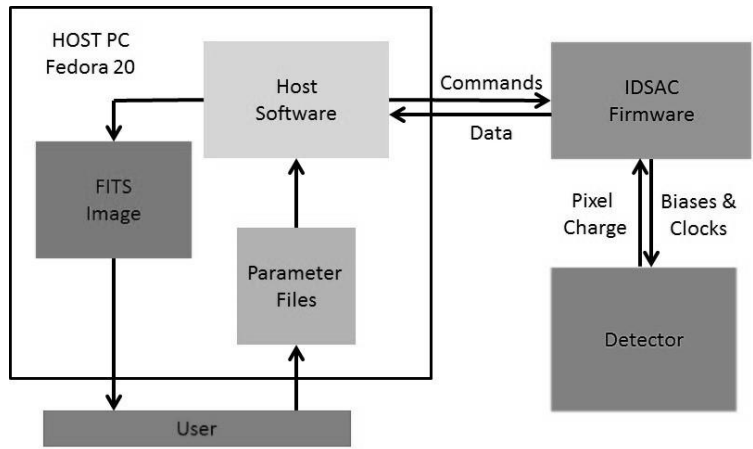

Fig 1 Schematic of IDSAC Architecture

Figure 2 shows the functional placements of components in IDSAC Single Board Controller (SBC). SBCs are available in two architecture versions. A back-plane based configuration allows multiple parallel SBCs in a single controller. The backplane card distributes power (supplied by Linear or Switched mode power supply) to all the SBCs. Except for the USB Interface to PC, all the controllers are interconnected via this board. Optional additional filtering for signal conditioning of all the Clocks and Biases connected to the CCD and Power supply input to the SBCs is also possible on the backplane card.

The SBC is also available in a standalone architecture (which has been used for all tests described in this paper) for use with single detectors or mosaics of small arrays as shown in Figure 2. Each SBC is designed to handle large detectors with up to four readout channels. The clock driver section is capable of driving 10 Bi-Level Serial Clocks for pixel transfers and 10 Multi-Level Parallel Clocks for line transfers with capacitive loads (about 330 pF and 50 nF for pixel and line clock respectively) at high speeds (e.g. 1 MHz for pixel clock). Virtex-5 series FPGA (Xilinx) is used as an embedded controller for generating readout clocks, reading four 16 bit serial ADCs and programming DACs for bias voltages. Each SBC can be connected to a host computer via USB2.0 connection with optional USB to Fiber converters. A newer version of IDSAC has USB 3.0 interface that can handle data rates up to 320 MBytes per second. More details about IDSAC will be provided in Burse et al. (in preparation).

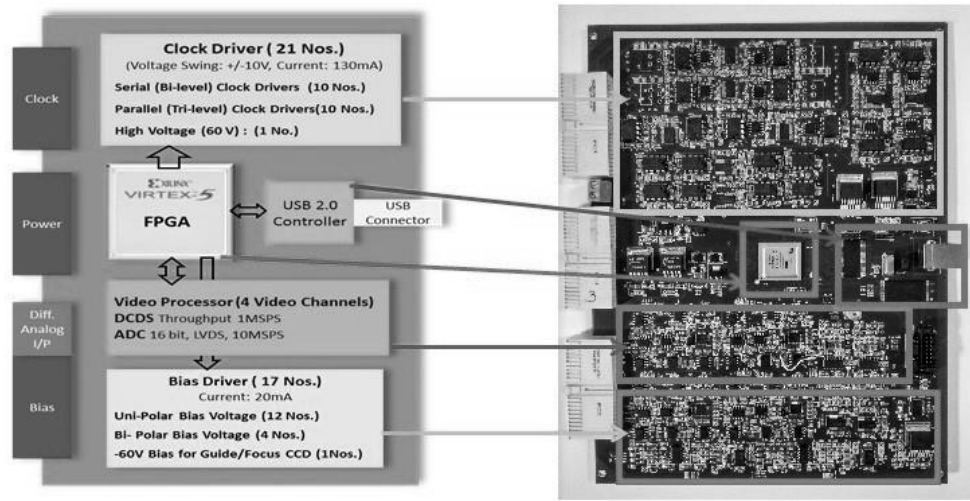

Fig 2 IDSAC SBC and it's block diagram

The Video Processor circuit which is shown in Figure 3 uses DCDS technique for processing CCD signal. The output of the CCD is fed to a low noise (white noise spectral density 7 nV/√Hz) Pre-Amplifier block via an AC coupler that provides input to the differential amplifier which also acts as ADC driver. A black level clamp is used in front of the preamplifier circuit but is not shown for the sake of simplicity and to focus on only the relevant components in the analytical model. Although the gain is variable, in this version of IDSAC the Pre-Amplifier and Differential Amplifier voltage gains are set to 2.5 and 4 respectively and it is used to match the ADC input range of -4.096 to +4.096 V. The ADC can sample this input at up to 10 Mega Samples per Second (MSPS) sampling rate and generate 16-bit data.

## 3 DCDS Analytical Noise Model

This section describes an analytical noise model of the IDSAC DCDS signal chain. A typical analog signal chain sends the detector output to the input of the Analog to Digital Converter (ADC) via multiple amplifier stages which set the gain. In IDSAC as shown in Figure 3, the CCD outputs are ac-coupled to the pre-amplifier inputs. The preamplifier output is then fed to a differential amplifier which is responsible for driving the ADC.

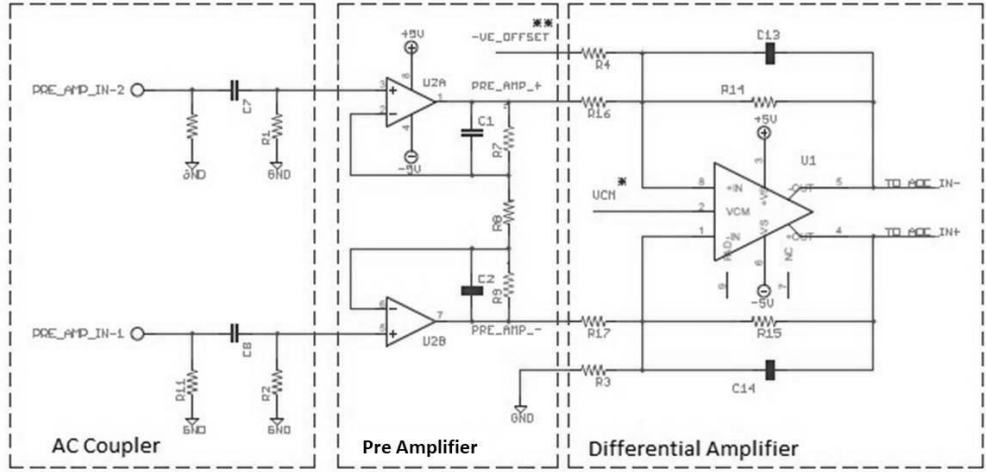

Fig 3 IDSAC Analog signal chain and its components.

For a generic signal chain consisting of n electronic blocks, $N_1$, $N_2$, ..., $N_n$ are the input referred noise powers of blocks 1, 2,..., $n$, and $G_1$, $G_2$,..., $G_n$ are their power gains, respectively. Based on the assumption that noise contribution of any block is independent of the other blocks in the signal chain, the total input referred noise power $N_{ip}$ can be defined as (15),

$$N_{ip} = N_1 + \frac{N_2}{G_1} + ... + \frac{N_n}{G_1 G_2 ... G_{n-1}}. \tag{1}$$

The AC coupler acts a high-pass filter to suppress the contribution from pink noise which dominates in the low-frequency domain. The IDSAC signal chain can handle single ended or differential output from CCD. The preamplifier block is chosen with high voltage gain (2.5) and low noise (7 nV/√Hz). One of the immediate applications of IDSAC is for the Devasthal Optical Telescope Integral Field Spectrograph (DOTIFS - 16). Each of the eight DOTIFS units uses a 2k×4k CCD which is to be read at 200 kilo Pixels Per Second (kPPS) per channel or faster. Driving the CCDs at this speed would require at least 5 MSPS ADC sampling rate to get ten samples per sampling intervals (reference and data level) including all the settling time (∼ 500 ns). The slew rate of the chosen amplifier for this stage is 180 V/μs. For a CCD output varying around 10 V, the bandwidth is set by the slew rate to ∼ 18 MHz. Such high bandwidth minimizes the settling time but introduces additional noise. So it is essential to add a bandwidth limiting circuit in the feedback path of the amplifier. This bandwidth is tunable and adjusted as per the analytical model prediction for noise optimization. Similarly, the ADC driver is chosen as a small noise contributor (2.25 nV/√Hz). For the scope of future improvement, we have opted for a 10 MSPS ADC with highly linear (±0.45 LSB INL) specification. The system performance parameters (gain, noise, readout speed, dynamic range, ADC data width etc.) of this version of IDSAC have been derived from the instrument requirements based on the science drivers.

The input referred noise power of each block has different components originating from various sources. We have investigated the effects of white, pink, bias, clock jitter, and Power Supply noise in detail in the subsections below.

*3.1 White Noise*

For any amplifier block in inverting configuration as shown in Figure 4, $R_1$, $R_2$, $R_3$ are the resistances in the input and feed back paths while $i_{nn}$, $i_{np}$ and $e_n$ are input referred current and voltage noises respectively. One can formulate output referred RMS noise voltage as an addition of the above mentioned white noise sources in quadrature integrated over frequency $f$ ( http://www.ti.com/lit/an/slva043b/slva043b.pdf ),

$$E_{\text{WN}} = \left[ \int \left( 4kTR_1 \left(\frac{R_2}{R_1}\right)^2 + 4kTR_2 + 4KTR_3 \left(\frac{R_1+R_2}{R_1}\right)^2 + e_n^2 \left(\frac{R_1+R_2}{R_1}\right)^2 + i_{\text{np}}^2 R_3^2 \left(\frac{R_1+R_2}{R_1}\right)^2 + i_{\text{nn}}^2 R_2^2 \right) \mathrm{d}f \right]^{1/2}, \quad (2)$$

where $k$ is Boltzmann constant, and $T$ is temperature. The first three terms signify the white noise contribution from three external resistances while the latter three terms provide how the amplifier internal voltage and current noises will contribute to the total output-referred noise.

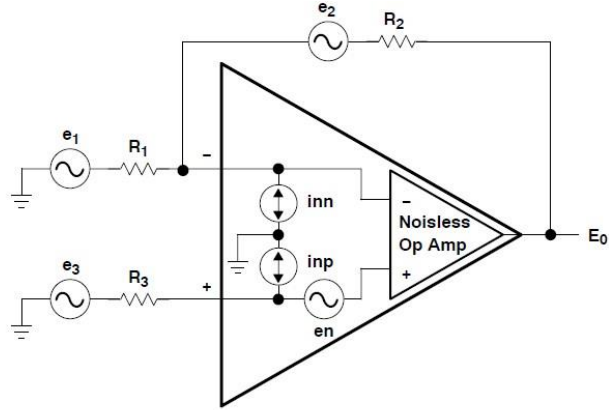

Fig 4 Amplifier Circuit Noise Sources taken from http://www.ti.com/lit/an/slva043b/slva043b. pdf.

Equation 2 shows the importance of bandwidth in white noise contribution of each stage. The bandwidth of the preamplifier stage should be large enough for the settling time and small enough to reduce the noise. Amplifier induced bandwidth of 18 MHz makes the settling time ($\sim$ 55 ns) small at the expense of significant white noise ($\sim$ 75 $\mu$V). On the other hand, a feedback bandwidth of 100KHz would provide minimal white noise ($\sim$ 6 $\mu$V) but high settling time ($\sim$ 10 $\mu$s) which would mean very slow readout speed that defies the advantages of DCDS. Using the theoretical model and the analytical tool, we explored the parameter space spanned by readout speed, number of samples and bandwidth to arrive at the optimal values for these design parameters. The veracity of the model/tool was validated by sampling the parameter space with actual measurements. 1.9 MHz is found to be the optimal bandwidth for this stage for the above-mentioned CCD throughput. Subsequent blocks require higher bandwidths (3 MHz for ADC driver input and 10 MHz for ADC input) to match up with the optimal bandwidth required to perform conversion at 5 MSPS.

Noise contribution of an ADC is defined by two quantities, Full-Scale noise power, and Signal to Noise Ratio (SNR - given at a particular bandwidth) both can be found in the ADC datasheet. So for a $-A$ Volt to $+A$ Volt peak to peak sine wave, the input referred noise power of an ADC block of $K$ SNR will be,

$$E_{WN-ADC} = \frac{A^2}{8K} \tag{3}$$

White noise contribution of different blocks in the signal chain predicted by the analytical method is given in table 1 for IDSAC DCDS signal chain shown in Figure 3 with bandwidth settings mentioned in the previous paragraph. The contribution of each block at its input and the input of the signal chain is mentioned separately to show the effect of the high gain input block in overall noise performance. For IDSAC, the analog signal chain white noise contribution is 27 $\mu$V. It is important to notice that use of sampling to reduce noise is still not taken into account up to this point. DCDS implemented with $p$ samples per sampling level (data and reference) would bring down the signal chain noise to $27\sqrt{(2/p)}$ $\mu$V. One can deduce that the minimum value of $p$ should be 3 to extract any advantage out of DCDS which is also demonstrated by measurements described in Section 4. This value is chosen because the total noise of a DCDS signal chain with $p < 3$ will be at least the white noise contribution of the signal chain if not higher. However, with a higher value of $p$ one can bring down the white noise contribution which is the dominating source of the total noise.

### 3.2 Pink Noise

The power spectral density (PSD) of pink noise varies in inverse proportion to the readout frequency, which suggests that this noise dominates below a corner frequency ($f = f_c$). For a given electronic block, let the white noise power be equal to the pink noise power at $f_c$. Then, $N_{pn} = N_{wn}f_c/f$, where $N_{pn}$ and $N_{wn}$ are the the pink noise and white noise power spectral densities respectively. One can integrate $N_{pn}$ over the relevant bandwidth to obtain the pink noise power of any block which is the square of the total RMS noise. The bandwidth is set by the lower cut-off frequency $f_L$ of the AC coupler and the upper cut-off is $f_c$ beyond which white noise dominates over the pink noise. Although this bandwidth estimation is true for the ideal case, for real scenario one has to incorporate the slow roll off of the filter transfer function. Hence, one can take $f_L/5$ and $5 \times f_c$ as the boundary frequencies, from which one can derive that the RMS pink noise voltage will be, $E_{pn} = \sqrt{N_{wn}f_c \ln(25f_c/f_L)}$ . Using the correct values of lower cut off and corner frequency it is found that, for the Pre-Amplifier block (used in DCDS signal chain in IDSAC), RMS pink noise voltage is less than 1 $\mu$V which means that its overall contribution is much smaller than white noise Power. Clearly, the effect of pink noise can be made negligible in systems where the operating bandwidth is much higher than the noise corner frequency $f_c$.

### 3.3 Power Supply Noise

The noise in the power supply lines can pass through to the output of the amplifier stage. If a power supply with RMS noise $\Delta V_s$ produces an RMS noise of $\Delta V_o$ at the output of an amplifier stage then the Power Supply Rejection Ratio (PSSR) is given by,

$$P = 20 \log \frac{\Delta V_s}{\Delta V_o.} \qquad (4)$$

To achieve 1 $\mu$V RMS power supply induced ripple at the output of an amplifier stage which has an 80dB PSRR at 100 kHz, use of a power supply with ≤ 10 mV RMS ripple is enough. Since PSRR decreases rapidly with the increase in frequency, it cannot be entirely trusted upon to keep power supply noise at acceptable levels. So additional filtering on board is provided to suppress the contribution of power supply noise. The RC circuit used in IDSAC sets the bandwidth of the decoupling circuit to 16 kHz. In this frequency range, the PSRR value is > 80 dB. In Section 4 we will show the results of power supply noise estimation by using two different types of sources.

*3.4 Bias Noise*

In IDSAC the biases and the clock rails that drives the CCD, are created by the FPGA using the regulated power supply available. Being the source, the power supply is the primary source of the noise present at the biases which contribute to the CCD output noise. As described in the previous subsection, we have suppressed the bias noise by applying additional filtering before the amplifier stage. In IDSAC, the power supply noise is found to be 7 mV RMS at 20 MHz power supply output bandwidth. For 500Hz filtering, the input bias noise becomes ∼ 2.3$\mu$V including noise from the amplifier (5.1 nV/ √Hz) and the digital to analog converter (75 nV/ √Hz) used for processing the bias with a voltage gain of 8.63. The bias noise is then modified by the CCD output electronics transfer function and gets added to the total electronics noise in quadrature. Assuming a maximum of 100% correlation between certain bias and the detector output, the CCD output noise (∼ 17 $\mu$V) found to be dominant than the bias noise.

*3.5 Clock Jitter Noise*

Jitter in the timing/clocking signals can introduce noise (charge induced) in CCD output and therefore needs to be modeled. As described in 17, the clock jitter noise $N_{CJ}$ can be calculated by,

$$N_{CJ} = [S + S_{OFF}][e^{-t_s/\tau_D} - e^{-t_s(1+J)/\tau_D}], \qquad (5)$$

where $S$ is the signal, $S_{OFF}$ is the offset (typically 100 mV), $t_s$ is the sampling time (∼ 200ns), $\tau_D$ is the CDS time constant (∼ 200ns) and $J$ is clock jitter. For crystal derived clocks as used in our system, $J$ is ∼ 2×10⁵ (20ppm) which amounts to hI∼7 $\mu$V in the absence of input signal and comparable to read noise. The jitter noise introduced by the clocks when added to CCD readout noise (16.65 $\mu$V) in quadrature increases the total output noise by 0.9 $\mu$V only. The effect of clock jitter is measured experimentally and discussed in Section 4.

*3.6 Analytical Noise Prediction*

The analytical noise model comprising of all the noise sources mentioned above have been incorporated into a python tool that predicts the noise performance of a signal chain when its components and working conditions are known. The user can vary different parameters for all the blocks (e.g., bandwidth, gain, slew rate, number of samples, sampling speed, etc.) to optimize the noise performance, which gives the freedom to narrow down the range of parameters without changing the hardware of the controller. This tool, as shown in Figure 5, can predict noise of a DCDS signal chain for given bandwidth, the number of samples in each sampling interval, CCD throughput and produces a plot of RMS read noise and samples per interval against the quantities mentioned above.

It can also provide comparisons of the performance of a DCDS signal chain with an "equivalent" analog CDS signal chain.

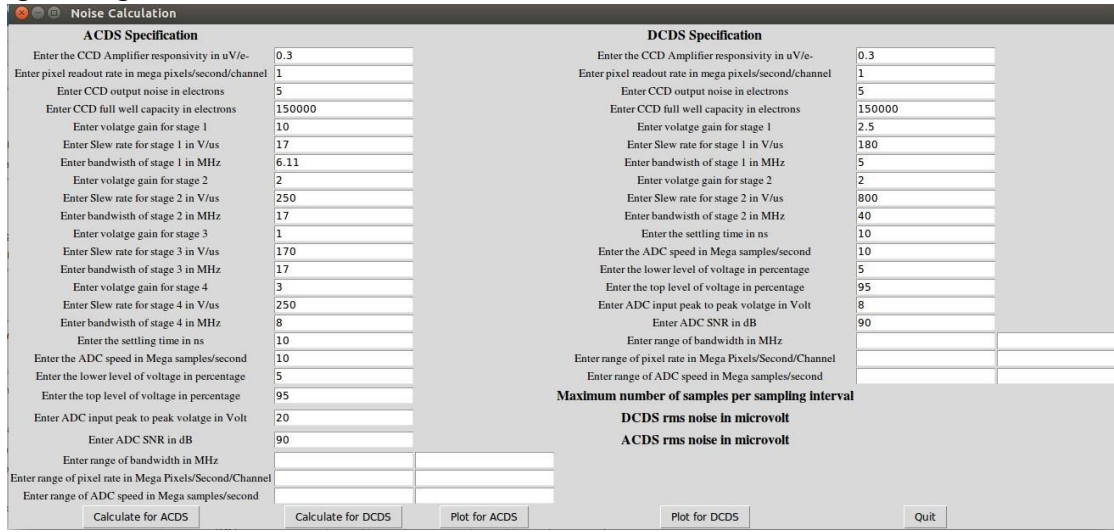

Fig 5 Analytical noise prediction tool. The tool can be used to predict noise of system with particular analog signal chain settings as well as trend of noise and number of samples against parameters like bandwidth, readout speed etc.

Figure 6 shows the predicted noise performance plotted against CCD readout rate when the maximum number of samples allowed by different system bandwidths, is used. If we increase the pixel readout rate, it will allow only fewer samples and in turn lead to higher noise. The effect of bandwidth and number of samples in the overall noise is interdependent and nonlinear. For a given readout rate, increasing the bandwidth will reduce settling time per sampling interval (reference or data level). However, this additional time should at least be equal to the sampling time (200 ns for IDSAC) to accommodate another sample for both the data and reference levels. If this time is less than required, then the noise will increase because the bandwidth increased without any increase in the number of samples. This effect can be seen for most pixel readout rates (> 150 kPPS) where an increase in bandwidth from 1 MHz to 2 MHz shows a jump in the maximum number of samples and as a result a decrease in RMS noise. This change in noise is because the settling time has changed from ∼ 900 ns (1 MHz) to ∼ 500 ns (2 MHz) and the time freed up is more than enough to accommodate another sample. However, increasing the bandwidth from 2 MHz (settling time ∼ 500 ns) to 5 MHz (settling time ∼ 350 ns) does not change the maximum number of samples, and hence the noise has increased. It is important to understand that the optimization of bandwidth and number of samples mostly depend on the sample-to-sample time and thus ADC speed. A faster ADC would provide smaller sampling time and in such a scenario the time gained from increasing the bandwidth (decreasing settling time) can be enough to raise the number of samples which would minimize overall noise.

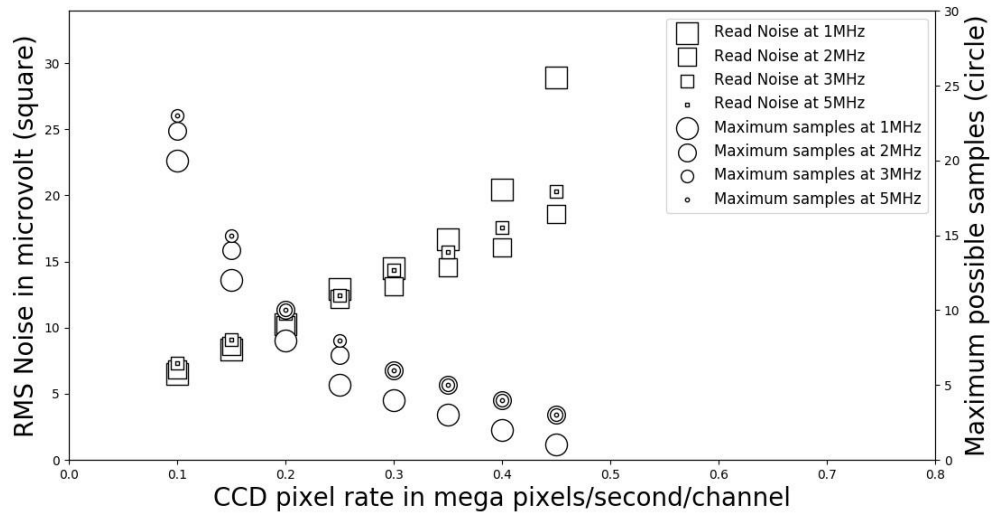

Fig 6 Plot shows the noise predicted by python tool using analytical model of DCDS signal chain of IDSAC for different bandwidth and CCD throughput, using 5 MSPS ADC sampling rate. The prediction of 23 $\mu$V, 20.2 $\mu$V, and 17.5 $\mu$V noise for 3, 4, and 5 samples matches with 23.8 $\mu$V, 21 $\mu$V, and 18.5 $\mu$V RMS noise respectively obtained from measurement for 350 kPPS readout at 1.9 MHz.

### 3.7 Simulation Study

Before proceeding to actual noise measurements with hardware, we also checked the IDSAC noise performance with a component level simulation using PSpice. The IDSAC analog signal chain was created as component level blocks by importing the noise models of each amplifier into the PSpice library. At this point, the input noise spectral density of each block is simulated (by removing the other blocks) to produce a plot of noise power vs. frequency (PSD). This PSD is then used to measure total noise within the bandwidth. In the next step, input noise of the entire analog chain is measured in a similar process. The simulation estimates 28 $\mu$V noise at the input of the signal chain while the analytical prediction is 27 $\mu$V. In the noise models of amplifiers used for simulation, filter transfer function has a more complicated roll off compared to the first order roll off assumed to build the analytical model. The simulation study shows the effect of all the noise components (from the amplifier noise model) and bandwidth in the overall noise but can not predict the effect of the number of samples.

### 4 Laboratory Noise Measurement

### 4.1 Setup

Since the 4K×2K CCDs for the DOTIFS (16) are under fabrication, the IDSAC characterization was carried out using a grade 5 e2v CCD42-40 (150,000 electrons full well capacity with 4.5 $\mu$V/electron amplifier responsivity) mounted in a cryogenic test Dewar as shown in Figure 7 in the test setup. The detector is mounted inside the liquid nitrogen cooled dewar. The dewar glass window is covered with an electromechanical shutter. The IDSAC and the CCD are connected via a flex cable. A USB cable is used to communicate with IDSAC from the software in the host computer. The CCD amplifier response is measured using an Fe 55 X-ray source (5.9 keV photons) which produces 1620 electrons per pixel. The X-ray source is mounted on the Dewar glass window and three images are obtained by exposing the CCD for 10 seconds each to the X-ray source. After dark subtraction, a histogram of the pixel counts is plotted for each channel of each image. The peak from K$\alpha$ (refer to Figure 8) is identified and fitted with a Gaussian to extract the mean value (~583 ADU) which in turn provides the gain of the channel in electron per Analog to Digital Unit (ADU). Since the ADU value in microvolts (12.5 $\mu$V/ADU) is known, the CCD amplifier responsivity is calculated to be 4.5 $\mu$V/electron.

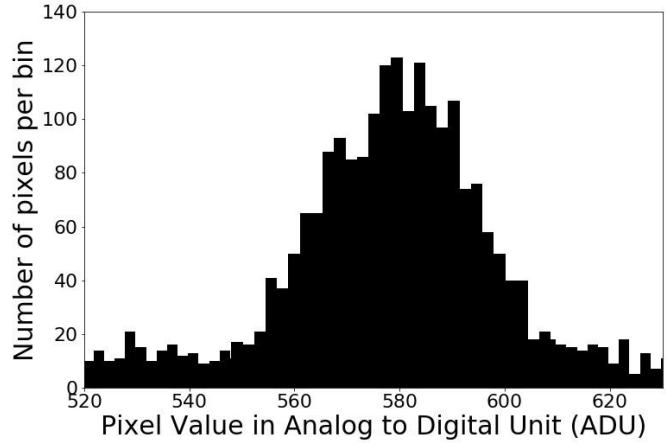

Fig 8 Part of the histogram of the pixel values showing the Fe 55 K$_\alpha$ peak.

The tests described in the following sections provide estimations of total noise contributed from all the relevant sources of noise. The analytical model and simulation tests described in Section 3 predict that the total noise of the system will be dominated by the white noise at the CCD operating speeds. Hence we have not used weighted averaging of the samples and this already gives a performance which is better than required for DOTIFS application.

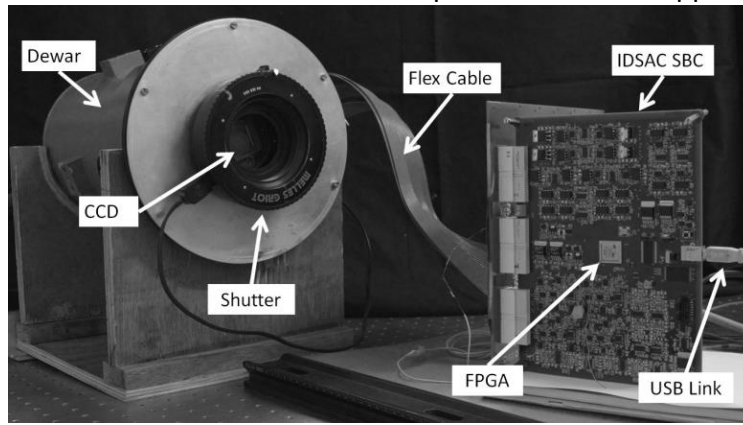

Fig 7 Test setup for IDSAC Characterization

*4.2 ADC Noise Measurement*

Our measurement tool for IDSAC characterization is the ADC in the analog signal chain. As the first step, the ADC performance was characterized by taking "images" with the inputs shorted to ground. The standard deviation of an area of 500 pixels×500 pixels of the resultant image (as shown in Figure 9) gives an estimate of the ADC RMS noise of IDSAC analog signal chain. This was found to be 1 electron RMS for 500 kPPS readout with three samples per interval at the input of the analog signal chain.

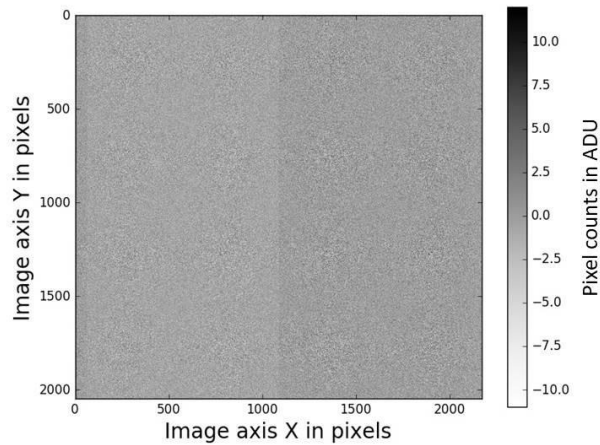

Fig 9 2048 pixels×2048 pixels image for characterizing ADC noise. ADC noise is measured by taking standard deviation of each channel of the image. Difference of gain of the two readout channel can be clearly visible in the image.

*4.3 Check for Missing Counts*

To check for any missing counts in the ADC output, exposures were taken with a slow 10 Hz triangle wave is fed to the input of the signal chain with sufficient amplitude to cover the full ADC input range (-4.096V to +4.096V). One can expect to have pixel counts to have all possible values from 0 to 65535 (the ADC output is 16 bit) since the triangle wave has much lower frequency than ADC readout speed. Figure 10 shows the histogram of the image shows a flat distribution showing no missing count.

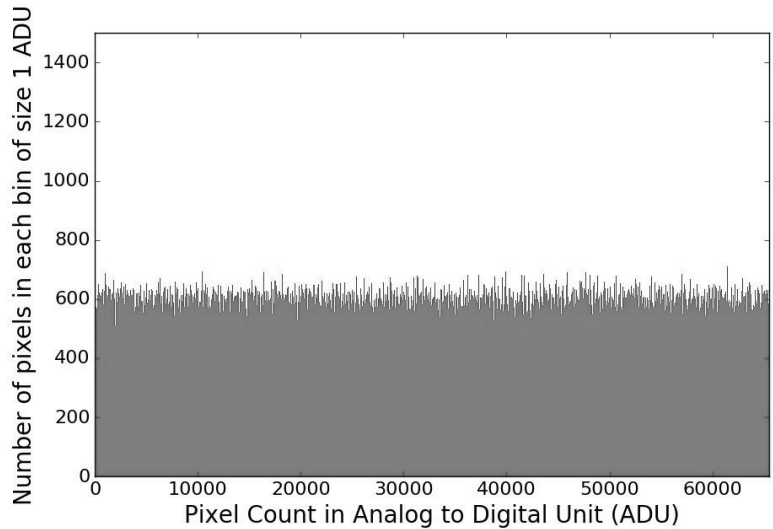

Fig 10 Histogram of pixel counts with bin size of one Analog to Digital Unit (ADU). The image is taken with a slow 10Hz triangle wave as the input to the signal chain.

*4.4 Bias and Clock Noise*

The next step is to understand the impact of the noise in the biases and clock rails. Each analog bias has been tested at a step of every 0.5 mV covering full range with appropriate resistor value to emulate the current drawn by the bias line and 200pF capacitance to mimic a typical gate.

- Bias voltage noise measurement is performed by feeding bias voltages to the input of the differential amplifier stage, and ADC output is grabbed as a full frame image. By varying the bias values, we changed the mean count in the frame as ∼ 8k, ∼ 32k and ∼ 60k to cover the entire ADC output range. For each value, four images are taken, and RMS read noise of an area of 500 pixels×500 pixels are measured. The bias noise is found to be 2.2 $\mu$V (0.5 electrons) which agrees with the theoretical prediction described in Subsection 3.4.

- All rails of the pixel and line clocks have similar noise to that of bias voltages.

- Apart from noise in the rails, rise and fall time of the clocks (in load condition) needs to be as per CCD requirement for proper charge transfer. For IDSAC, Rise and Fall time of the pixel clocks, measured with 330pF capacitive load (to mimic CCD loads as mentioned in the datasheet) is found to be ∼ 50 ns RMS as shown in Figure 11. The rise and fall time can be changed from parameter file but are set to 50 ns based on the requirement mentioned in the CCD data sheet.

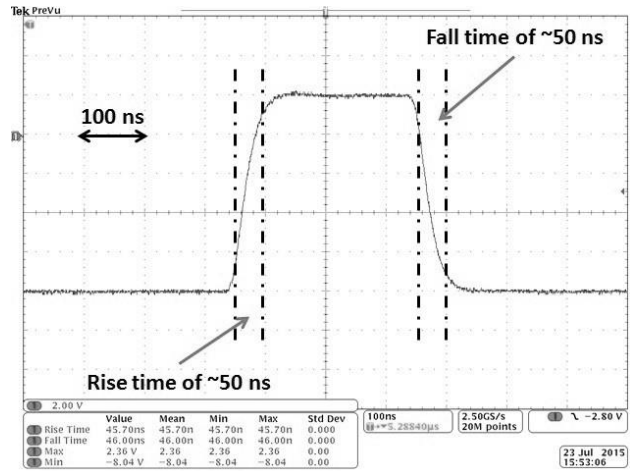

Fig 11 Digital oscilloscope screen grab showing Rise and Fall time of a typical pixel clock with 330pF capacitive load to mimic CCD capacitance.

*4.5 Controller Read Noise*

The read noise of analog chain is measured by analyzing 500 pixels×500 pixels region of 5 full frames taken with both preamplifier inputs shorted to ground. Table 2 shows that the RMS noise is found to be 5.3 electrons (23.8 $\mu$V) for a CDS throughput of 350 kPPS and three samples in each sampling interval as shown in Figure 12. The measurement matches closely with the prediction of the analytical model of 5.1 electrons (23 $\mu$V) RMS noise for the same configuration. The model prediction is also very close to actual measurements with four samples (4.7 electrons (21.2 $\mu$V) measured vs. 4.5 electrons (20.25 $\mu$V) predicted) and five samples (4.1 electrons (18.5 $\mu$V) measured vs. 3.9 electrons(17.6 $\mu$V) predicted). The analytical and measured noise values are slightly different as the model does not accommodate the nonlinear noise response of FET switches. Also as mentioned in Subsection 3.7, the assumption of 1st order amplifier response roll-off is another reason for the discrepancies. As expected both the model and the measurement show that an increase in the number of samples provides a decrease in RMS noise for same CCD throughput.

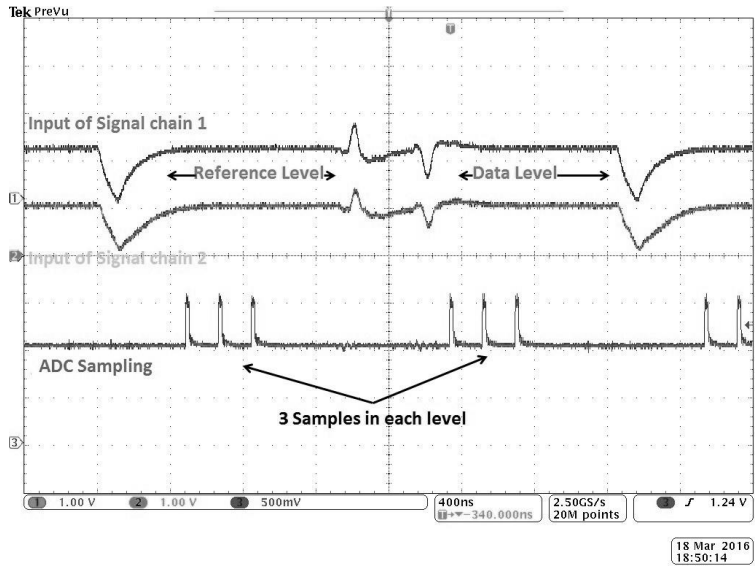

Fig 12 Digital oscilloscope screen grab showing 3 samples for reference and data level.

Once the characterization of the entire analog signal chain is performed, we characterized CCD and signal chain together. At the input of the signal chain, CCD readout noise and signal chain noise get added in quadrature. As described in Subsection 4.1, a grade 5 e2v CCD42-40 is read by the CDS signal chain. 7.5 electrons (33.75 $\mu$V), 7.1 electrons (32 $\mu$V), and 6.7 (30.15 $\mu$V) electrons RMS noise is measured for CCD and signal chain together with 3, 4, and 5 samples respectively.

Table 3 shows that the total RMS noise (CCD and readout electronics together) $\sim$ 10% higher in the right channel compared to that in the left channel for 200 kPPS throughput and eight samples per sampling interval, irrespective of whether the CCD is readout in single output mode or dual output mode. Since both the signal chains showed 2.6 electrons (11.7 $\mu$V) RMS noise with grounded inputs, the discrepancy is ascribed to different gains of the two CCD output amplifiers. Here onward we have described all the results based on the experiments performed with the left channel.

In order to get the best noise performance at a given CCD readout rate, the bandwidth of the signal chain has to appropriately optimized. The bandwidth of the signal chain was optimized for the highest possible CCD readout rate of 500 kPPS for IDSAC system. The RMS read noise of the signal chain was measured for different bandwidth settings of the second stage (differential amplifier) while keeping the bandwidths of the other stages fixed. The measured RMS noise of 4.8 electrons (21.6 $\mu$V) is very close to the model predicted noise of 4.7 electrons (21.1 $\mu$V) for 7 MHz bandwidth. The model also predicts the noise for other bandwidth values with similar accuracy. Table 4 shows that 3 MHz is the optimum bandwidth for this readout rate with three samples. This result shows that not only signal chain input bandwidth but also bandwidth of other stages affect the overall noise.

To understand the trend of noise performance of CCD and IDSAC analog signal chain against Readout Speed, the RMS readout noise of CCD and controller was measured for 100, 200 and 350 kPPS per channel CCD throughput. During measurement, the number of samples was also varied, and the measured RMS noise is compared to the prediction from the model as shown in Table 5.

As described earlier, for each readout speed, the noise goes down with an increase in the number of samples. Since CCD readout noise increases with readout rate, for the same number of samples combined noise of CCD and controller also increases.

In order to eliminate the possibility of charge induced noise in the measurements, the readout noise of IDSAC was measured while reverse clocking and compared to that of normal clocking. In standard clocking, the clocking pattern will shift the charge towards the CCD output through which CCD is read, but in the case of reverse clocking, the charge gets transferred in the opposite direction of the output channel. CCD noise is calculated by subtracting electronics noise from the overall noise in quadrature. The comparison provided in table 6 shows that noise introduced by charges is negligible and matches with analytical prediction as described in Subsection 3.5.

*4.6 Power Supply Noise*

Linear mode power supplies usually show less than 10 mV RMS ripple. Hence they are often used in spite of being bulky in size. IDSAC uses a switched mode power supply board (SMPS) which makes it very compact and lightweight. At the output of the power supply, a capacitor combination of 180 and 0.1 $\mu$F is used to provide filtering. The CCD is read at a speed of 200 kPPS with 8 samples per interval. IDSAC delivered a ripple of 4.9 electrons (22 $\mu$V) RMS when powered by the SMPS board as compared to 4.7 electrons (21.2 $\mu$V) RMS when supplied by a conventional linear power supply.

*4.7 Noise Analysis in Fourier Domain*

Fourier analysis of an image brings out any repeated fixed pattern noise in the detector or readout system. For this purpose, a bias image obtained at 200 kPPS (with 8 samples per level) is converted into a 1D array by appending lines one after another. A separate 1D timing array is also created which contains timing of each pixel reading. Fast Fourier Transform (FFT) is performed on the 1D data array, and a sideband is plotted against frequency as shown in figure 13. Typically fixed pattern noise shows up as peaks at multiple frequencies. Thus any static, repetitive pattern on the detector would show up on the plot as a combination of those multiple frequency peaks. We have calibrated the FFT peaks by injecting additional known signals (amplitude in electrons) at different known frequencies. Using the relation of the FFT magnitudes of these injected peaks with their time domain amplitudes in electrons, we have calibrated the FFT magnitude for all the necessary frequencies which shows peaks significantly higher than the noise floor. The dominant peak in IDSAC data is at 0Hz which corresponds to the mean value (∼1000 electrons) of the 1D array. Other frequency peaks are at least three orders of magnitude weaker than the 0Hz peak and the integrated RMS noise voltage is equivalent to peak to peak amplitude of ∼1 electron (4.5 $\mu$V). Figure 13(d) shows a second strongest peak at ∼67 kHz which corresponds to a pattern of 3 pixels length with an amplitude of 1.6 $\mu$V RMS. Compared to the total noise of 13.5 $\mu$V RMS, the noise associated with this peak is negligible when considered in quadrature. We do not find peaks at the same or corresponding (scaled by pixel rate) frequencies in FFT of bias images read at 100, 350 or 500 kPPS as shown in Figure 13 (a), (c), and (d) respectively.

## 5 Summary

IDSAC's versatility stems from the separation of the core CCD control process of bias and clock generation (done in the FPGA firmware) from configuration settings through text-based parameter files residing in the PC. A Single Board Controller (SBC) along with the SMPS board can drive four analog signal channels simultaneously at speeds up to 500 kPPS per channel for capacitive loads of about 330 pF and 50 nF for pixel and line clocks respectively. DCDS leads to a compact analog processing chain and flexibility to accommodate a variety of noise/performance conditions. We have characterized and optimized the DCDS signal chain in IDSAC through an analytical model, software simulation, and laboratory measurements. The prediction from analytical model matches with simulation and experimental results within 10% accuracy. We have developed and verified the theoretical understanding of a DCDS signal chain and built a tool to do aid optimization of design choices so as to quickly arrive at a solution that meets the performance requirements of a specific instrument.

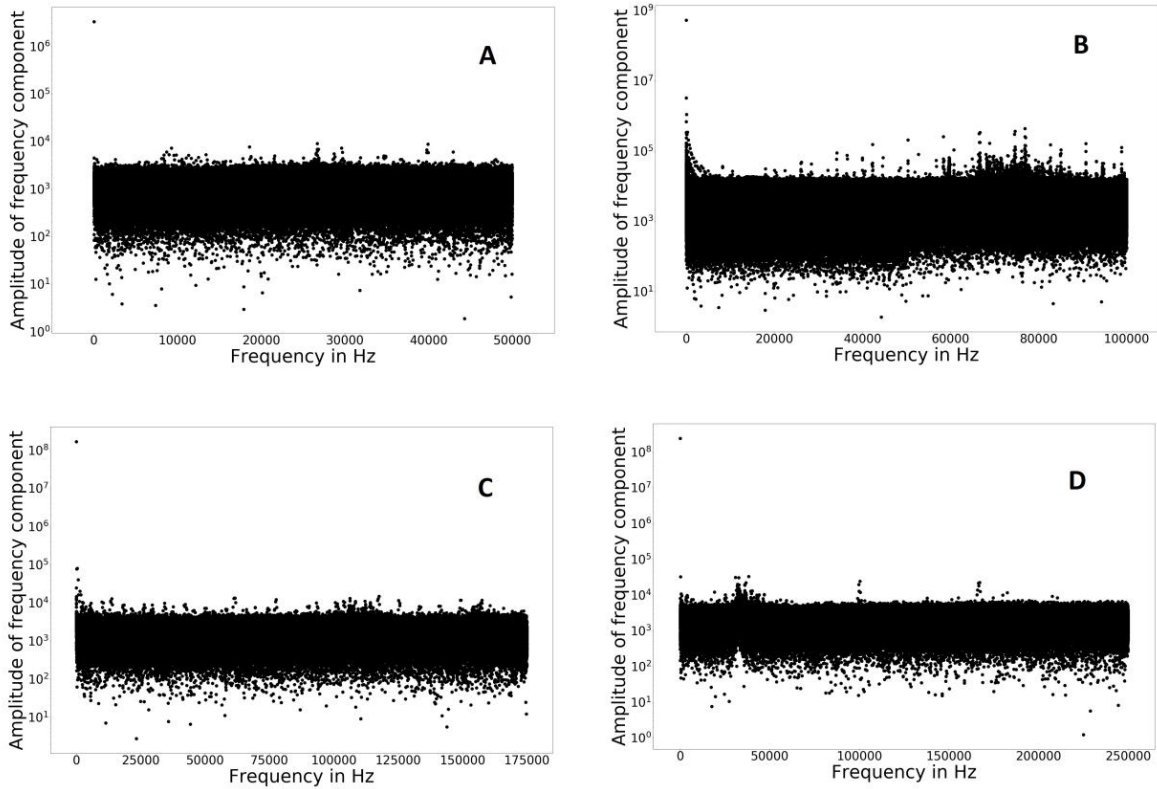

Fig 13 Fourier transforms of 1D arrays converted from bias images obtained at 100 (A), 200 (B), 350 (C) and 500 (D) kPPS from single CCD output.

Optimization of the number of samples and bandwidth is carried out for best noise performance for a particular CCD readout rate. For 500 kPPS CCD throughput measured at 5 MSPS, RMS read noise of 3.8 electrons (17 $\mu$V) at 3 MHz system bandwidth is the minimum provided the number of samples per interval is fixed at 3. At these CCD readout speeds, white noise is found to be the dominant contributor. We achieve the required noise performance for DOTIFS with simple averaging of samples. However, in situations where other noise sources (eg: pink noise) dominate, a weighted averaging scheme may fare better. The controller architecture allows such weighted averaging. We have found that 1.9 MHz and 3 MHz are the optimum input bandwidth for IDSAC for 200 kPPS and 350 kPPS CCD readout speed respectively. A Fourier transform analysis shows that the fixed pattern noise is very small. We obtain 2.6 electrons (11.7 $\mu$V) RMS noise using 8 samples per interval at 200 kPPS CCD throughput for the optimized analog signal chain which is lower than the CCD noise of 3.7 electrons (16.65 $\mu$V). The readout noise performances of IDSAC signal chain at 100, 200, and 500 kPPS are found to be 7, 11.7, and 17 $\mu$V respectively. For a DCDS chain developed by Clapp (11) they report a total RMS read noise for the above CCD throughputs are 9, 13, and 23.5 $\mu$V respectively for a total of 60 samples in a pixel readout period. Gach et al. (10) have obtained ~8 $\mu$V controller noise by reading out EEV CCD 42-20 at 100 kPPS with a total of 100 samples. This shows the DCDS implemented in IDSAC provides similar noise characteristics to its predecessors however with a slightly better optimization which has originated from the theoretical understanding of the noise components. Overall IDSAC is easily optimizable and adaptable to the needs for controlling most of the CCDs used in optical astronomy. Although IDSAC has been characterized with an E2V CCD 42-40, but it meets all the requirements of controlling DOTIFS E2V CCD 44-82 in terms of readout speed (200 kPPS), acceptable read noise (18 $\mu$V RMS), rise and fall time of clocks (50 ns), data digitization (16 bit), support of single-ended or differential readout mode etc.

*Acknowledgments*

This research was partially supported by Council of Scientific and Industrial Research, Human Resource Development Group, Govt. of India (http://www.csirhrdg.res.in/). We also thank the referees for their valuable comments and suggestions.

Table 1 Theoretical prediction of white noise power contribution from different stages of IDSAC DCDS signal chain. The contribution of each block at its input and at the input of the signal chain is mentioned separately to show the effect of high gain input block in overall noise performance.

| DCDS stage | RMS Noise Voltage in $\mu$V at block input | RMS Noise Voltage in $\mu$V at signal chain input | Voltage Gain | Bandwidth in MHz |
|---|---|---|---|---|
| Pre-Amplier | 26.5 | 26.5 | 2.5 | 1.9 |
| ADC Driver | 4.8 | 1.9 | 4 | 3 |
| ADC | 22 | 2.2 | NA | 10 |

Table 2 Effect of number of samples in measured Read noise. The test is performed at 350 kPPS readout rate and 1.9 MHz system bandwidth (14).

| Number of Samples in data and reference Level | 3 Samples | 4 Samples | 5 Samples |
|---|---|---|---|
| RMS Noise of Analog Signal Chain only in electrons ($\mu$V) | 5.3 (23.85) | 4.7 (21.15) | 4.1 (18.5) |
| RMS Noise of Analog Signal chain and CCD combined in electrons ($\mu$V) | 7.5 (33.75) | 7.1 (32) | 6.7 (30.15) |

Table 3 RMS noise performance of the two analog signal chains of IDSAC used to readout CCD for 200 kPPS CCD throughput with 8 samples per sampling interval.

| Analog signal chain used for | Readout Mode Single Channel | Readout mode Dual Channel |
|---|---|---|
| RMS Noise of left signal chain and the left channel of CCD in electrons($\mu$V) | 4.55 (20.5) | 4.55 (20.5) |
| RMS Noise of right signal chain and the right channel of CCD in electrons ($\mu$V) | 5.02 (22.6) | 5.04 (22.7) |

Table 4 Effect of bandwidth in RMS readout noise of analog signal chain with and without CCD. The test is performed at 500 kPPS readout rate and 3 samples per sampling interval by varying the differential amplifier stage bandwidth (14). The input bandwidth of the pre-amplifier stage was fixed at 1.9 MHz.

| DCDS System Bandwidth in MHz | 7 MHz | 4.8 MHz | 3 MHz | 1.9 MHz |
|---|---|---|---|---|
| RMS Noise of Analog Signal Chain only in electrons ($\mu$V) | 4.8 (21.6) | 4.4 (19.6) | 3.8 (17.1) | 4 (18) |
| RMS Noise of Analog Signal Chain and CCD combined in electrons ($\mu$V) | 7.7 (34.65) | 7.4 (33.3) | 7.09 (31.9) | 7.2 (32.4) |

Table 5 Noise comparison of CCD and IDSAC signal chain at different readout speed with varying number of samples per interval with 1.9 MHz and 3 MHz bandwidth for pre-amplifier and differential amplifier stage respectively.

| No. of Samples | Overall Noise in Electrons ($\mu$V) of CCD and controller | | | | | |
| --- | --- | --- | --- | --- | --- | --- |
| | Readout Speeds | | | | | |
| | 100 kPPS | | 200 kPPS | | 350 kPPS | |
| | Measured | Predicted | Measured | Predicted | Measured | Predicted |
| 3 | 6.58 (29.6) | 6.20 (27.9) | 6.65 (29.9) | 6.20 (27.9) | 7.5 (33.8) | 7.25 (32.6) |
| 4 | 6.09 (27.4) | 5.69 (25.6) | 6.16 (27.7) | 5.63 (25.3) | 7.1 (32.0) | 6.80 (30.6) |
| 5 | 5.55 (25.0) | 5.37 (24.2) | 5.57 (25.0) | 5.30 (23.9) | 6.7 (30.2) | 6.55 (29.5) |
| 6 | 5.09 (22.9) | 5.14 (23.1) | 5.08 (22.9) | 5.06 (22.8) | | |
| 7 | 4.90 (22) | 4.97 (22.4) | 4.87 (21.9) | 4.89 (22.0) | | |
| 8 | 4.62 (20.8) | 4.76 (21.4) | 4.55 (20.5) | 4.70 (21.1) | | |
| 9 | 4.38 (19.1) | 4.65 (20.9) | | | | |
| 10 | 4.25 (19.6) | 4.57 (20.6) | | | | |
| 11 | 4.08 (18.4) | 4.49 (20.2) | | | | |
| 12 | 4.00 (18.0) | 4.43 (19.9) | | | | |

Table 6 Readout noise comparison of CCD and controller for normal and reverse clocking at 200 kPPS CCD throughput and 8 samples per sampling interval

| Readout noise measured at 200 kPPS, 8 samples per interval | Normal clocking noise in elecrons ($\mu$V) | Reverse clocking noise in electrons ($\mu$V) |
| --- | --- | --- |
| CCD and Controller (measured) | 4.5 (20.25) | 4.6 (20.7) |
| Controller only (measured) | 2.6 (11.7) | 2.6 (11.7) |
| CCD only (calculated) | 3.7 (16.65) | 3.8 (17.1) |